\documentclass[11pt]{article}
\usepackage{amsmath, amssymb, amscd, epsfig}

\newtheorem{thm}{Theorem}
\newtheorem{prop}[thm]{Proposition}
\newtheorem{lem}[thm]{Lemma}

\newtheorem{cor}[thm]{Corollary}

\renewcommand{\epsilon}{\varepsilon}
\renewcommand{\phi}{\varphi}

\newcommand{\BB}{\mathbb}

\newcommand{\pf}{\noindent {\it Proof. }}
\newcommand{\qed}{\nopagebreak $\qquad$ $\square$ \vskip5pt}

\newcommand{\sgn}{\operatorname{sign}}

\begin{document}

\title{\bf Hyperbolic Cauchy Integral Formula for the Split Complex Numbers}
\author{Matvei Libine}
\maketitle

\begin{abstract}
In our joint papers \cite{FL1,FL2} we revive quaternionic
analysis and show deep relations between quaternionic analysis,
representation theory and four-dimensional physics.
As a guiding principle we use representation theory of various real forms
of the conformal group. We demonstrate that the requirement of unitarity
of representations naturally leads us to the extensions of the Cauchy-Fueter
and Poisson formulas to the Minkowski space, which can be viewed as
another real form of quaternions.
However, the Minkowski space formulation also brings some technical
difficulties related to the fact that the singularities of the kernels in
these integral formulas are now concentrated on the light cone instead of
just a single point in the initial quaternionic picture.
But the same phenomenon occurs when one passes from the complex numbers
to the split complex numbers (or hyperbolic algebra).
So, as a warm-up example we proved an analogue of the Cauchy integral formula
for the split complex numbers.
On the other hand, there seems to be sufficient interest in such formula
among physicists. For example, see \cite{KS} and the references therein.
\end{abstract}

In this short article we give a Cauchy-type integral formula for solutions of
the wave equation $\square_{1,1} F =0$ on
$\BB R^2 = \{ (x,y); \: x, y \in \BB R \}$, where
$$
\square_{1,1} F = \frac {\partial^2}{\partial x^2} F
- \frac {\partial^2}{\partial y^2} F.
$$
More precisely, we write elements $(x,y)$ of $\BB R^2$ as $Z = x + jy$ and
equip $\BB R^2$ with multiplication operation so that $j^2 = 1$.
Then $\BB R^2$ becomes an algebra over $\BB R$, and we denote this algebra
by $\BB R^{1,1}$.
We introduce two differential operators
$$
\partial_{1,1} = \frac 12 \Bigl(
\frac{\partial}{\partial x} + j \frac{\partial}{\partial y} \Bigr)
\qquad \text{and} \qquad
\partial_{1,1}^+ = \frac 12 \Bigl(
\frac{\partial}{\partial x} - j \frac{\partial}{\partial y} \Bigr).
$$
Note that
$$
\partial_{1,1} \partial_{1,1}^+ = \partial_{1,1}^+ \partial_{1,1} =
\frac 14 \square_{1,1}.
$$
We will give an integral formula for differentiable functions
$F: \BB R^{1,1} \to \BB R^{1,1}$ such that $\partial_{1,1}^+ F =0$.

\begin{prop}
Let $f, g: \BB R \to \BB R$ be a pair of smooth single-variable functions.
Then
\begin{equation}  \label{sol}
F(x,y) = \bigl(f(x+y) + jf(x+y)\bigr) + \bigl(g(x-y) - jg(x-y)\bigr)
\end{equation}
is a solution of $\partial_{1,1}^+ F =0$. Moreover, any smooth solution of
$\partial_{1,1}^+ F =0$ is of this form.
\end{prop}

\pf
The first statement is straightforward.
To prove the second statement, let $F$ be such that $\partial_{1,1}^+ F =0$.
Write
$$
F= \frac{1+j}2 \cdot F + \frac{1-j}2 \cdot F.
$$
Then
$$
0 = \frac{1+j}2 \cdot \partial_{1,1}^+ F = 
\frac{1+j}2 \cdot \frac 12 \Bigl(
\frac{\partial}{\partial x} - j \frac{\partial}{\partial y} \Bigr) F
= \frac 12 \Bigl(
\frac{\partial}{\partial x} - \frac{\partial}{\partial y} \Bigr)
\Bigl(\frac{1+j}2 \cdot F \Bigr)
$$
which implies that $\frac{1+j}2 \cdot F$ can be written as 
$f(x+y) + jf(x+y)$ for some smooth single-variable function
$f: \BB R \to \BB R$.
Similarly,
$$
0 = \frac{1-j}2 \cdot \partial_{1,1}^+ F = 
\frac{1-j}2 \cdot \frac 12 \Bigl(
\frac{\partial}{\partial x} - j \frac{\partial}{\partial y} \Bigr) F
= \frac 12 \Bigl(
\frac{\partial}{\partial x} + \frac{\partial}{\partial y} \Bigr)
\Bigl(\frac{1-j}2 \cdot F \Bigr)
$$
which implies that $\frac{1-j}2 \cdot F$ can be written as 
$g(x-y) - jg(x-y)$ for some smooth single-variable function
$g: \BB R \to \BB R$.
\qed

\begin{cor}  \label{constant}
Let $F: \BB R^{1,1} \to \BB R^{1,1}$ be a smooth function satisfying
$\partial_{1,1}^+ F =0$, then $\frac{1+j}2 \cdot F$ is constant along the lines
$x+y=\operatorname{const}$ and $\frac{1-j}2 \cdot F$ is constant along the
lines $x-y=\operatorname{const}$.
\end{cor}

\begin{lem}
If $F, G: \BB R^{1,1} \to \BB R^{1,1}$ are smooth functions satisfying
$\partial_{1,1}^+ F = \partial_{1,1}^+ G =0$, then so is their product $FG$.
\end{lem}

\pf
$\partial_{1,1}^+ (FG) = (\partial_{1,1}^+ F)G + F(\partial_{1,1}^+ G)=0$.
\qed

\begin{lem}
Let $dZ = dx + jdy$, and let $F: \BB R^{1,1} \to \BB R^{1,1}$ be a
differentiable function. Then
$$
d(F\,dZ)= 2j \cdot (\partial_{1,1}^+ F) \,dx \wedge dy.
$$
In particular,
$$
d(F\,dZ)=0 \qquad \Longleftrightarrow \qquad \partial_{1,1}^+ F =0.
$$
\end{lem}

Let $U \subset \BB R^{1,1}$ be an open subset with piecewise smooth boundary
$\partial U$. We give a canonical orientation to $\partial U$ as follows.
The positive orientation of $U$ is determined by $\{1,j\}$.
Pick a smooth point $p \in \partial U$ and let $\overrightarrow{n_p}$
be a non-zero vector in $T_p \BB R^{1,1}$ perpendicular to
$T_p\partial U$ and pointing outside of $U$.
We give $\partial U$ an orientation so that a tangent vector
$\overrightarrow{\tau_p} \ne 0$ at $p \in \partial U$ points in the
positive direction if and only if
$\{\overrightarrow{n_p}, \overrightarrow{\tau_p}\}$ in $\BB R^{1,1}$.

\begin{cor}  \label{green}
Let $U \subset \BB R^{1,1}$ be an open bounded subset as above,
let $V$ be an open neighborhood of the closure $\overline{U}$
And let $F: V \to \BB R^{1,1}$ be a smooth function such that
$\partial_{1,1}^+ F =0$. Then
$$
\int_{\partial U} F\,dZ =0.
$$
\end{cor}

For $Z = x + jy$ set
$$
Z^+ = x - jy, \qquad N(Z)= x^2 -y^2, \qquad \|Z\| = \sqrt{x^2+y^2}.
$$

\begin{lem}
\begin{enumerate}
\item
An element  $Z = x + jy \in \BB R^{1,1}$ is invertible if and only if
$N(Z)=x^2 - y^2 \ne 0$; in which case $Z^{-1} = \frac{Z^+}{N^2(Z)}$;
\item
If $Z,W \in \BB R^{1,1}$, then $N(ZW)=N(Z) \cdot N(W)$;
\item
The function $K(Z)= Z^{-1} = \frac{Z^+}{N^2(Z)}$ satisfies
$\partial_{1,1}^+ K =0$ wherever $N(Z) \ne 0$.
\end{enumerate}
\end{lem}

Observe that
\begin{multline}
K(x+jy) = \frac{x-jy}{x^2-y^2}
= \frac{1+j}2 \cdot \frac{x-jy}{x^2-y^2} +
\frac{1-j}2 \cdot \frac{x-jy}{x^2-y^2}  \\
= \frac{1+j}2 \cdot \frac{x-y}{x^2-y^2} +
\frac{1-j}2 \cdot \frac{x+y}{x^2-y^2}
= \frac{1+j}2 \cdot \frac1{x+y} +
\frac{1-j}2 \cdot \frac1{x-y}.
\end{multline}
Let $(\BB R^{1,1})^{\times} = \{x+jy \in \BB R^{1,1} ;\: x^2-y^2 \ne 0\}$
be the set of invertible elements.
For $\epsilon \in \BB R$, define
$K_{\epsilon}: (\BB R^{1,1})^{\times} \to \BB R^{1,1} \otimes_{\BB R} \BB C$
by:
$$
K_{\epsilon}(x+jy) = \frac{1+j}2 \cdot \frac1{x+y +i\epsilon\cdot\sgn(x-y)} +
\frac{1-j}2 \cdot \frac1{x-y + i\epsilon\cdot\sgn(x+y)}.
$$
Then on each connected component of $(\BB R^{1,1})^{\times}$ the function
$K_{\epsilon}$ is of the type (\ref{sol}), hence satisfies
$\partial_{1,1}^+ K_{\epsilon} =0$.

\begin{prop}[Integral Formula]
Let $R>0$, set $U = \{ Z \in \BB R^{1,1} ;\: |N(Z)| < R \}$ and let
$V$ be an open neighborhood of the closure $\overline{U}$.
Suppose $F: V \to \BB R^{1,1}$ is a smooth bounded function satisfying
$\partial_{1,1}^+ F =0$ and pick any $Z_0=x_0+jy_0 \in U$, then
\begin{equation}  \label{main11}
f(Z_0) = \frac 1{2\pi i} \lim_{\epsilon \to 0^+} \int_{\{|N(Z)|=R\}}
K_{Z_0,\epsilon}(Z) \cdot F(Z) \,dZ,
\end{equation}
where
$$
K_{Z_0,\epsilon} =
\frac{1+j}2 \cdot \frac1{x-x_0+y-y_0 +i\epsilon\cdot\sgn(x-y)} +
\frac{1-j}2 \cdot \frac1{x-x_0-y+y_0 + i\epsilon\cdot\sgn(x+y)},
$$
the hyperbolas $N(Z)= \pm R$ are oriented as $\partial U$
(i.e. counterclockwise) and the improper integral is defined as
$$
\int_{\{|N(Z)|=R\}} K_{Z_0,\epsilon}(Z) \cdot F(Z) \,dZ =
\lim_{S \to \infty} \int_{\{|N(Z)|=R\} \cap \{\|Z\| \le S\}}
K_{Z_0,\epsilon}(Z) \cdot F(Z) \,dZ.
$$
\end{prop}

Note that $K_{Z_0,0} = K(Z-Z_0)$, so we can regard $K_{Z_0,\epsilon}(Z)$
as a perturbation of $K(Z-Z_0)= \frac 1{Z-Z_0}$.
Thus the integral formula formally looks identical to the
Cauchy integral formula for holomorphic functions.

\pf
Write $K_{Z_0,\epsilon}$ as $K_{Z_0,\epsilon}^+ + K_{Z_0,\epsilon}^-$,
where
\begin{align*}
K_{Z_0,\epsilon}^+(Z) &= \frac{1+j}2 \cdot
\frac1{x-x_0+y-y_0 +i\epsilon\cdot\sgn(x-y)}  \\
\text{and} \qquad
K_{Z_0,\epsilon}^-(Z) &= \frac{1+j}2 \cdot
\frac1{x-x_0-y+y_0 +i\epsilon\cdot\sgn(x+y)}.
\end{align*}
To prove (\ref{main11}) it is enough to show that
\begin{equation}  \label{Kpm}
2\pi i \cdot \frac{1 \pm j}2 \cdot F(Z_0) =
\lim_{\epsilon \to 0^+} \biggl(
\lim_{S \to \infty} \int_{\{|N(Z)|=R\} \cap \{\|Z\| \le S\}}
K_{Z_0,\epsilon}^{\pm}(Z) \cdot F(Z) \,dZ \biggr).
\end{equation}
Note that both $K_{Z_0,\epsilon}^{\pm}$ satisfy 
$\partial_{1,1}^+ K_{Z_0,\epsilon}^{\pm} =0$
and that the integrand is a closed form.
We also observe that
\begin{align*}
(1+j) \,dZ &= (1+j)(dx+jdy) = (1+j)d(x+y)  \\
\qquad \text{and} \qquad (1-j) \,dZ &= (1-j)(dx+jdy) = (1-j)d(x-y).
\end{align*}
We change coordinates to $(u,v)$ so that
$$
u=x+y, \qquad v=x-y, \qquad u_0=x_0+y_0, \qquad v_0=x_0-y_0.
$$
Then
\begin{align*}
K_{Z_0,\epsilon}^+(Z) \cdot F(Z) \,dZ &= 
\frac{1+j}2 \cdot
\frac {F(u,v)}{u-u_0 +i\epsilon\cdot\sgn(v)} \,du,  \\
K_{Z_0,\epsilon}^-(Z) \cdot F(Z) \,dZ &= 
\frac{1-j}2 \cdot
\frac {F(u,v)}{v-v_0 +i\epsilon\cdot\sgn(u)} \,dv.
\end{align*}
The hyperbolas $\{N(Z)=\pm R\}$ in these coordinates become
$\{uv=\pm R\}$.

Let
$$
U_S = U \cap \{|u|,|v| < S \} =
\{Z \in \BB R^{1,1} ;\: -R < N(Z) < R \text{ and } |u|,|v| < S \},
$$
and orient its boundary $\partial U_S$ as in Corollary \ref{green}, i.e.
counterclockwise.

\begin{lem}
$$
\lim_{S \to \infty} \biggl( \int_{\{|N(Z)|=R\} \cap \{|u|,|v| \le S\}}
K_{Z_0,\epsilon}^{\pm}(Z) \cdot F(Z) \,dZ
- \int_{\partial U_S} K_{Z_0,\epsilon}^{\pm}(Z) \cdot F(Z) \,dZ \biggr) =0
$$
\end{lem}

\pf
The difference of integrals in question is integral of
$K_{Z_0,\epsilon}^{\pm}(Z) \cdot F(Z) \,dZ$ over the four straight segments
of the boundary $\partial U_S$:
$$
\{Z \in \BB R^{1,1} ;\: |v|=S \text{ and } |u| \le R/S \}
\qquad \text{and} \qquad
\{Z \in \BB R^{1,1} ;\: |u|=S \text{ and } |v| \le R/S \}.
$$
The length of each segment is $2R/S$.
Since $F$ is bounded and $|K_{Z_0,\epsilon}^{\pm}| \le \frac 1{\epsilon}$,
the integrals over these segments tend to zero as $S \to \infty$.
\qed

Since the form $K_{Z_0,\epsilon}^+(Z) \cdot F(Z) \,dZ$ is closed, by Stokes'
theorem applied to the two regions $U_S \cap \{v>0\}$ and $U_S \cap \{v<0\}$
we have
\begin{multline*}
\int_{\partial U_S} K_{Z_0,\epsilon}^+(Z) \cdot F(Z) \,dZ  \\
= \frac{1+j}2 \int_{u=-S}^{u=S} \biggl(
\frac {F(u,v)}{u-u_0 -i\epsilon} \biggr|_{v=0}
- \frac {F(u,v)}{u-u_0 +i\epsilon} \biggr|_{v=0} \biggr) \,du  \\
= \frac{1+j}2 \int_{u=-S}^{u=S}
\frac {2i\epsilon}{(u-u_0)^2 +\epsilon^2} \cdot F(u,0)\,du.
\end{multline*}
As $S\to \infty$, we get
$$
\lim_{S \to \infty}
\int_{\partial U_S} K_{Z_0,\epsilon}^+(Z) \cdot F(Z) \,dZ
= \frac{1+j}2 \int_{-\infty}^{\infty}
\frac {2i\epsilon}{(u-u_0)^2 +\epsilon^2} \cdot F(u,0) \,du.
$$
Finally,
\begin{multline*}
\lim_{\epsilon \to 0^+} \int_{-\infty}^{\infty}
\frac {2i\epsilon}{(u-u_0)^2 +\epsilon^2} \cdot F(u,0) \,du =
\lim_{\epsilon \to 0^+} \int_{-\infty}^{\infty}
\frac {2i}{(\frac{u-u_0}{\epsilon})^2 +1} \cdot F(u,0)
\,d \Bigl(\frac{u-u_0}{\epsilon}\Bigr)  \\
= \lim_{\epsilon \to 0^+} \int_{-\infty}^{\infty}
\frac {2i}{t^2 +1} \cdot F(u_0 + \epsilon t,0) \,d t = 2\pi i \cdot F(u_0,0).
\end{multline*}
By Corollary \ref{constant},
$\frac{1+j}2 \cdot F(u_0,0) = \frac{1+j}2 \cdot F(u_0,v_0)$.
This proves (\ref{Kpm}) for $K_{Z_0,\epsilon}^+(Z)$.

Similarly, the form $K_{Z_0,\epsilon}^-(Z) \cdot F(Z) \,dZ$ is closed,
by Stokes' theorem applied to the two regions $U_S \cap \{u>0\}$ and
$U_S \cap \{u<0\}$ we have
\begin{multline*}
\int_{\partial U_S} K_{Z_0,\epsilon}^-(Z) \cdot F(Z) \,dZ  \\
= \frac{1-j}2 \int_{v=-S}^{v=S} \biggl(
\frac {F(u,v)}{v-v_0 -i\epsilon} \biggr|_{u=0}
- \frac {F(u,v)}{v-v_0 +i\epsilon} \biggr|_{u=0} \biggr) \,dv  \\
= \frac{1-j}2 \int_{u=-S}^{u=S}
\frac {2i\epsilon}{(v-v_0)^2 +\epsilon^2} \cdot F(0,v)\,dv.
\end{multline*}
As $S\to \infty$, we get
$$
\lim_{S \to \infty}
\int_{\partial U_S} K_{Z_0,\epsilon}^-(Z) \cdot F(Z) \,dZ
= \frac{1-j}2 \int_{-\infty}^{\infty}
\frac {2i\epsilon}{(v-v_0)^2 +\epsilon^2} \cdot F(0,v) \,dv.
$$
Finally,
\begin{multline*}
\lim_{\epsilon \to 0^+} \int_{-\infty}^{\infty}
\frac {2i\epsilon}{(v-v_0)^2 +\epsilon^2} \cdot F(0,v) \,dv =
\lim_{\epsilon \to 0^+} \int_{-\infty}^{\infty}
\frac {2i}{(\frac{v-v_0}{\epsilon})^2 +1} \cdot F(0,v)
\,d \Bigl(\frac{v-v_0}{\epsilon}\Bigr)  \\
= \lim_{\epsilon \to 0^+} \int_{-\infty}^{\infty}
\frac {2i}{t^2 +1} \cdot F(0,v_0 + \epsilon t) \,d t = 2\pi i \cdot F(0,v_0).
\end{multline*}
By Corollary \ref{constant},
$\frac{1-j}2 \cdot F(0,v_0) = \frac{1-j}2 \cdot F(u_0,v_0)$.
This proves the second part of (\ref{Kpm}).
\qed

Next we show how the requirement in the integral formula that $F$
is bounded can be dropped.

\begin{cor}
As before, let $R>0$, set $U = \{ Z \in \BB R^{1,1} ;\: |N(Z)| < R \}$ and let
$V$ be an open neighborhood of the closure $\overline{U}$.
Suppose $F: V \to \BB R^{1,1}$ is a smooth function satisfying
$\partial_{1,1}^+ F =0$ and pick any $Z_0=x_0+jy_0 \in U$.
Let $\phi: \BB R \to [0,1]$ be any smooth function with compact support
such that $\phi(0)=1$. Then
$$
f(Z_0) = \frac 1{2\pi i} \lim_{\epsilon \to 0^+} \int_{\{|N(Z)|=R\}}
K_{Z_0,\phi,\epsilon}(Z) \cdot F(Z) \,dZ,
$$
where
$$
K_{Z_0,\phi,\epsilon} =
\frac{1+j}2 \cdot
\frac {\phi(x-x_0+y-y_0)}{x-x_0+y-y_0 +i\epsilon\cdot\sgn(x-y)} +
\frac{1-j}2 \cdot
\frac {\phi(x-x_0-y+y_0)}{x-x_0-y+y_0 + i\epsilon\cdot\sgn(x+y)}.
$$
\end{cor}

\pf
The functions
\begin{align*}
F_1(Z) &= \frac{1+j}2 \cdot \phi(x-x_0+y-y_0) \cdot F(Z-Z_0)  \\
\text{and} \qquad
F_2(Z) &= \frac{1-j}2 \cdot \phi(x-x_0-y+y_0) \cdot F(Z-Z_0)
\end{align*}
are bounded, satisfy $\partial_{1,1}^+ F_1 = \partial_{1,1}^+ F_2 = 0$ and
$$
F_1(Z_0) = \frac{1+j}2 \cdot F(Z_0),
\qquad
F_2(Z_0) = \frac{1-j}2 \cdot F(Z_0).
$$
Hence the integral formula (\ref{main11}) applies and the result follows.
\qed


\begin{thebibliography} {[FL2]}
\bibitem[FL1]{FL1} I.~Frenkel, M.~Libine, {\em Quaternionic analysis,
representation theory and physics}, arXiv:0711.2699, submitted.
\bibitem[FL2]{FL2} I.~Frenkel, M.~Libine,
{\em Split quaternionic analysis and the separation of the series for
$SL(2,\mathbb R)$}, work in progress.
\bibitem[KS]{KS} A.~Khrennikov, G.~Segre,
{\em An Introduction to Hyperbolic Analysis}, arXiv:math-ph/0507053.
\end{thebibliography}
\end{document}